\documentclass[twocolumn,preprintnumbers,amsmath,nofootinbib,amssymb]{revtex4}
\usepackage{amssymb}
\usepackage{latexsym}
\usepackage{color}
\usepackage{enumerate}
\usepackage{graphicx}
\usepackage{hyperref}
\usepackage{color}

\def\be{\begin{equation}}
\def\ee{\end{equation}}
\def\bea{\begin{eqnarray}}
\def\eea{\end{eqnarray}}
\begin{document}
\title{The generalized and extended uncertainty principles and their implications on the Jeans mass}
\author{$^1$H. Moradpour\footnote{h.moradpour@riaam.ac.ir}, $^1$A. H. Ziaie\footnote{ah.ziaie@riaam.ac.ir}, $^1$S. Ghaffari\footnote{sh.ghaffari@riaam.ac.ir} and $^2$F. Feleppa \footnote{feleppa.fabiano@gmail.com}}
\address{$^1$ Research Institute for Astronomy and Astrophysics of Maragha (RIAAM), P.O. Box
55134-441, Maragha, Iran}
\address{$^2$ Department of Physics, University of Trieste, via Valerio
	2, 34127 Trieste, Italy}
\begin{abstract}
The generalized and extended uncertainty principles
affect the Newtonian gravity and also the geometry of the
thermodynamic phase space. Under the influence of the latter, the
energy-temperature relation of ideal gas may change. Moreover, it
seems that the Newtonian gravity is modified in the framework of
the R\'{e}nyi entropy formalism motivated by both the long-range
nature of gravity, and the extended uncertainty principle. Here,
the consequences of employing the generalized and extended
uncertainty principles, instead of the Heisenberg uncertainty
principle, on the Jeans mass are studied. The results of working
in the R\'{e}nyi entropy formalism are also addressed. It is shown
that unlike the extended uncertainty principle and the R\'{e}nyi
entropy formalism which lead to the same increase in the Jeans
mass, the generalized uncertainty principle can decrease it. The
latter means that a cloud with mass smaller than the standard
Jeans mass, obtained in the framework of the Newtonian gravity,
may also undergo the gravitational collapse process.
\end{abstract}

\maketitle
\section{Introduction}
The Heisenberg uncertainty principle (HUP) is in full agreement
with the Bekenstein entropy (BE) which can produce the Newtonian
gravity (NG) (Ali \& Tawfik 2013; Ali 2014; Awada \& Ali 2014a;
Awada \& Ali 2014b; Majumder 2011; Srednicki 1993; Verlinde 2011;
Wang et al. 2009). NG has a crucial role in getting some
well-known criteria appeared in study the structure formation
problems such as those of Chandrasekhar and Jeans (Longair 1998).
In this regard, it has been shown that the generalized and
extended uncertainty principles, namely GUP and EUP, respectively,
modify NG and can also affect the Chandrasekhar limit (Ong 2018;
Ong \& Yao 2018). Indeed, GUP and EUP change the Bekenstein limit
of entropy (Ali \& Tawfik 2013; Ali 2014; Awada \& Ali 2014a;
Awada \& Ali 2014b; Majumder 2011; Moradpour et al. 2019; Wang et
al. 2009), and even, EUP gives a motivation to use the R\'{e}nyi
entropy formalism instead of the Boltzmann-Gibbs (BG) entropy to
study the gravitational systems (Moradpour et al. 2019).

The quantum gravity scenarios generally propose (Bolan \& Cavaglia
2005; Bambi \& Urban 2008; Park 2008; Kempf et al. 1995; Nozari \&
Fazlpour 2006)
\begin{eqnarray}\label{gup}
\Delta x\Delta p\geq\frac{\hbar}{2}[1+\beta(\Delta
x)^2+\eta(\Delta p)^2+\gamma],
\end{eqnarray}

\noindent where $\beta$, $\eta$ and $\gamma$ are positive and
$\gamma$ can depend on the expectation values of $p$ and $x$.
Smallest uncertainties are obtained for $\gamma=0$ (Nozari \&
Fazlpour 2006), and non-zero minimal uncertainty in both position
and momentum is guaranteed whenever $\beta$ and $\eta$ are
positive (Kempf et al. 1995; Nozari \& Fazlpour 2006). Throughout
this paper, we set $\gamma=0$ (Nozari \& Fazlpour 2006), and
although some authors call Eq.~(\ref{gup}) as GUP (Kempf et al.
1995; Nozari \& Fazlpour 2006), by following Refs.~(Ali \& Tawfik
2013; Ali 2014; Awada \& Ali 2014a; Awada \& Ali 2014b; Majumder
2011; Moradpour et al. 2019; Wang et al. 2009) and~(Bolan \&
Cavaglia 2005; Bambi \& Urban 2008; Park 2008), we call the
$\beta=0$ and $\eta=0$ cases as GUP and EUP, respectively. It has
been shown that the accelerated universe and the galaxies flat
rotation curves may be modelled within the frameworks of GUP and
EUP without considering any odd energy source, namely dark energy
and dark matter (Awada \& Ali 2014b; Moradpour et al 2017;
Moradpour et al 2018a; Moradpour et al. 2019b; Komatsu 2017), a
result also signalling us to a deep connection between the fine
and large structures. Finally, it is worthwhile mentioning that
$\beta$ may also reveal the non-extensive features of
gravitational systems (Moradpour et al. 2019), and Eq.~(\ref{gup})
affects the thermodynamics of early universe (Nozari \& Fazlpour
2006).

The gravitational potential energy of a cloud with mass $M$ and
radius $R$, formed due to gravitational potential $V(r)$, is
evaluated as

\begin{eqnarray}\label{gpe}
U=\int_0^MV(r)dM.
\end{eqnarray}

\noindent %where
%$\rho(R)\big(\equiv\frac{M}{\frac{4\pi}{3}R^3}\big)$ is the cloud
%density.
In the framework of NG, where $V_N(r)=-G\frac{M}{r}$, the
gravitational potential energy ($U_N$) is obtained as

\begin{eqnarray}\label{3}
U_N=-\int_0^R\frac{GM}{r}4\pi\rho(r)r^2dr=-\frac{3GM^2}{5R},
\end{eqnarray}

\noindent written by assuming that the cloud density is constant
(or equally $\rho(r)=\rho_0\equiv const$). In the framework of BG
entropy, as a special case of Shannon entropy, the kinetic energy
($K$) of a cloud, approximated by an ideal gas composed of $N$
non-interacting particles with mass $\mu$ (and thus $M=N\mu$) at
temperature $T$, is

\begin{eqnarray}\label{virial}
\big<K\big>=\frac{3}{2}NK_BT,
\end{eqnarray}

\noindent where $K_B$ denotes the Boltzmann constant. Based on the
Virial theorem, the cloud is in Virial equilibrium if
$\big<K\big>=-\frac{1}{2}U$, and gravitational collapse may happen
if $\big<K\big><-\frac{1}{2}U$ leading to

\begin{eqnarray}\label{virial1}
NK_BT<\frac{GM^2}{5R},
\end{eqnarray}

\noindent for NG. Using the
$R=(\frac{3M}{4\pi\rho_0})^{\frac{1}{3}}$ relation in rewriting
Eq.~(\ref{virial1}), one reaches

\begin{eqnarray}\label{virial2}
M>M^J\equiv(\frac{5K_BT}{G\mu})^{\frac{3}{2}}(\frac{3}{4\pi\rho_0})^{\frac{1}{2}},
\end{eqnarray}

\noindent as the lower bound of the cloud mass to collapse. Any
change in the Jeans mass affects the fragmentation of cloud
(Forgan \& Rice 2011), and in fact, the predictions about the
structure formation at different scales (Capozziello et al 2012;
Roshan \& Abbassi 2014; Vainio \& Vilja 2016; de Martino \&
Capolupo 2017). As an example, while the mass of CB $188$, as a
Bok globule, is smaller than $M^J$, it has a protostar (Kandori et
al 2005; Vainio \& Vilja 2016) meaning that new physics, such as
modified gravity, is unavoidable to decrease the Jeans mass
(Vainio \& Vilja 2016).

As we saw, NG and the ordinary extensive statistical mechanics,
based on Shannon entropy, form the backbone of the obtained $M^J$,
and on the other, Newtonian force is the direct result of using BE
allowed by HUP (Ali \& Tawfik 2013; Ali 2014; Awada \& Ali 2014a;
Awada \& Ali 2014b; Majumder 2011; Moradpour et al 2019; Srednicki
1993; Verlinde 2011; Wang et al. 2009). In fact, NG is the weak
field limit of general relativity, and the modified versions of
general relativity, such as the $f(R)$ (Capozziello et al 2012),
scalar-tensor-vector (Roshan \& Abbassi 2014) and
Eddington-inspired Born-Infield gravities (de Martino \& Capolupo
2017), may decrease the Jeans mass by increasing the Newtonian
force at the weak field limit. Additionally, it also seems that
the consequences of non-zero values of $\beta$ and $\eta$ are not
limited to the fine structure, and indeed, their effects can even
be seen at the cosmic and galaxies scales (Awada \& Ali 2014b;
Komatsu 2017; Moradpour et al 2017; Moradpour et al 2018a;
Moradpour et al 2019; Moradpour et al. 2019b; Nozari \& Fazlpour
2006), and the physics of black hole (Moradpour et al 2019; Ong
2018; Ong \& Yao 2018), white dwarf and neutron stars (Ong 2018;
Ong \& Yao 2018).

Moreover, it has been claimed that the generalized entropies, such
as the R\'{e}nyi entropy, should be used instead of the Shannon
entropy to study the systems including long-range interactions
such as gravity (R\'{e}nyi 1970; Masi 2005; Moradpour et al 2017;
Moradpour et al 2018a; Moradpour et al. 2019b; Komatsu 2017). In
fact, even the ideal gas can give us permission to use the
generalized entropies such as those of the R\'{e}nyi and Tsallis
instead of the Shannon entropy (Bir\'{o} 2013). The Jeans mass
changes whenever instead of the BG entropy, Tsallis entropy is
considered to determine the thermodynamic quantities of system
(Jiulin 2004; Lima et al 2002). As we previously mentioned, it
also seems that a deep connection between the R\'{e}nyi statistics
and the fine structure parameters such as $\beta$, appeared in
EUP, may exist (Moradpour et al. 2019). The above arguments
motivate us to study the Jeans mass in the R\'{e}nyi entropy
formalism.

In the next section, we firstly study the effects of EUP on $M^J$.
Additionally, motivated by the fact that there is a deep
connection between EUP and R\'{e}nyi statistics (Moradpour et al.
2019), the implications of the R\'{e}nyi entropy on the Jeans mass
are also addressed in Sec.~($\textmd{II}$). The consequences of
GUP on $M^J$ are studied in third section, where it is also shown
that GUP may help us in explaining the existence of star in some
Bok globules. The last section is devoted to a summary. We also
follow the $c=\hbar=G=K_B=1$ units.
%%%%%%%%%%%%%%%%%%%%%%%%%%%%%%%%%%%%%%%%%%%%%%%%%%%%%%%%%%%%%%%%%%%%%
\section{EUP and Jeans instability: the role of R\'{e}nyi entropy}

Whenever $\eta=0$, EUP is obtained from Eq.~(\ref{gup}) as (Bolan
\& Cavaglia 2005; Bambi \& Urban 2008; Park 2008; Kempf et al.
1995; Ong \& Yao 2018; Wang et al. 2009)

\begin{eqnarray}\label{EUP1}
\Delta x\Delta p\geq\frac{1}{2}[1+\beta(\Delta x)^2],
\end{eqnarray}

\noindent which modifies the BE entropy ($\frac{A}{4}$) as
(Moradpour et al 2019)

\begin{eqnarray}\label{ent1}
S=\frac{\pi}{4\beta}\ln(1+\beta\frac{A}{\pi}),
\end{eqnarray}

\noindent where $A=4\pi r^2$ denotes the system boundary located
at radius $r$. Clearly, BE is recovered at the $\beta=0$ limit
(Moradpour et al 2019). It is also useful to mention that,
mathematically, the above entropy is called the R\'{e}nyi entropy
(Moradpour et al 2019).

%namely R\'{e}nyi entropy, where $A=4\pi r^2$ denotes the system
%boundary located at radius $r$. Clearly, BE is recovered at the
%appropriate limit $\delta=0$ (Moradpour et al 2019). Since the $\delta$
%parameter and $s$ are also motivated by the long-range nature of
%gravity (Komatsu 2017; Moradpour et al 2017; Moradpour et al 2018a; Moradpour et al
%2018b; Moradpour et al. 2019b), our analysis in this section can
%also be interpreted as the effects of employing the R\'{e}nyi
%statistical mechanics instead of the Boltzmann-Gibbs entropy
%formalism on the value of Jeans mass.

Using the entropic force approach, one can reach (Moradpour et al
2018b)

\begin{eqnarray}\label{f1}
F_S=-\frac{Mm}{r^2}\frac{1}{4\beta r^2+1},
\end{eqnarray}

\noindent instead of NG ($F_N=-\frac{Mm}{r^2}$). In agreement with
the asymptotic behavior of $S$, it is also obvious that NG is
recovered for $\beta=0$. Bearing the $F=-m\frac{dV}{dr}$ relation
in mind, the modified Newtonian potential $V_s$ is obtained as

\begin{eqnarray}\label{p1}
&&V_S(r)=-\frac{1}{m}\int
F_sdr\\&&=-M[\frac{1}{r}+2\sqrt{\beta}\big(\tan^{-1}(2\sqrt{\beta}r)-\frac{\pi}{2}\big)],\nonumber
\end{eqnarray}

\noindent where the integration constant has been chosen to meet
the $V_S(r\rightarrow\infty)=0$ condition. Now, applying this
potential to a cloud with mass $M$ and density $\rho_0$, and by
following the recipe led to Eq.~(\ref{3}), we get

\begin{eqnarray}\label{gpe1}
&&\!\!\!\!\!\!\!U_S=\int_0^RV_S(r)dM=-\frac{3M^2}{5R}+2\sqrt{\beta}U_c=U_N+2\sqrt{\beta}U_c,\nonumber\\
&&\!\!\!\!\!\!\!U_c=-\int_0^RM\big(\tan^{-1}(2\sqrt{\beta}r)-\frac{\pi}{2}\big)dM=U_NRD\nonumber,\\
&&\!\!\!\!\!\!\!D=\frac{5}{6}\bigg(\frac{(1+\xi^6)\tan^{-1}(\xi)}{\xi^6}+\frac{5\xi^2-3\xi^4-15}{15\xi^5}-\frac{\pi}{2}\bigg),\nonumber\\
&&\!\!\!\!\!\!\!\xi=2\sqrt{\beta}R,
\end{eqnarray}

\noindent for the corresponding gravitational potential energy
which can briefly be written as $U_S=U_N(1+\xi D)$.

The Helmholtz free energy ($A$) of a monatomic ideal gas which
consists of $N$ particles with mass $\mu$ at temperature $T$ and
satisfies EUP~(\ref{EUP1}) is obtained as (Chung \& Hassanabadi
2019)

\begin{eqnarray}\label{hf1}
&&A=-T\bigg[N\ln\big(4\pi(2\mu\pi T)^{\frac{3}{2}}
V_{eff}\big)-\ln(N!)\bigg],
\end{eqnarray}

\noindent where

\begin{eqnarray}\label{vef}
&&V_{eff}=\frac{4\pi}{\beta}\bigg[(\frac{3V}{4\pi})^{\frac{1}{3}}-\frac{1}{\beta^{\frac{1}{2}}}\tan^{-1}\big(\sqrt{\beta}(\frac{3V}{4\pi})^{\frac{1}{3}}\big)\bigg],
\end{eqnarray}

\noindent in which $V$ denotes the container volume (Chung \&
Hassanabadi 2019). Using $P=-(\frac{\partial
	A}{\partial{V}})_{T,N}$, one can see the ideal gas law ($PV=NT$)
is modified in this situation (Chung \& Hassanabadi 2019). Now,
following the $S_g=-(\frac{\partial A}{\partial{T}})_{V,N}$ and
$K=A+TS_g$ relations, one reaches

\begin{eqnarray}\label{vir1}
&&K=\frac{3}{2}NT,
\end{eqnarray}

\noindent for the internal energy of ideal gas nothing but
Eq.~(\ref{virial}). Here, $S_g$ represents also the gas entropy
which differs from that of the system boundary~(\ref{ent1}) (Ali
\& Moussa 2014; Moradpour et al 2019; Moradpour et al. 2014).

\begin{figure}
	\begin{center}
		\includegraphics[scale=0.28]{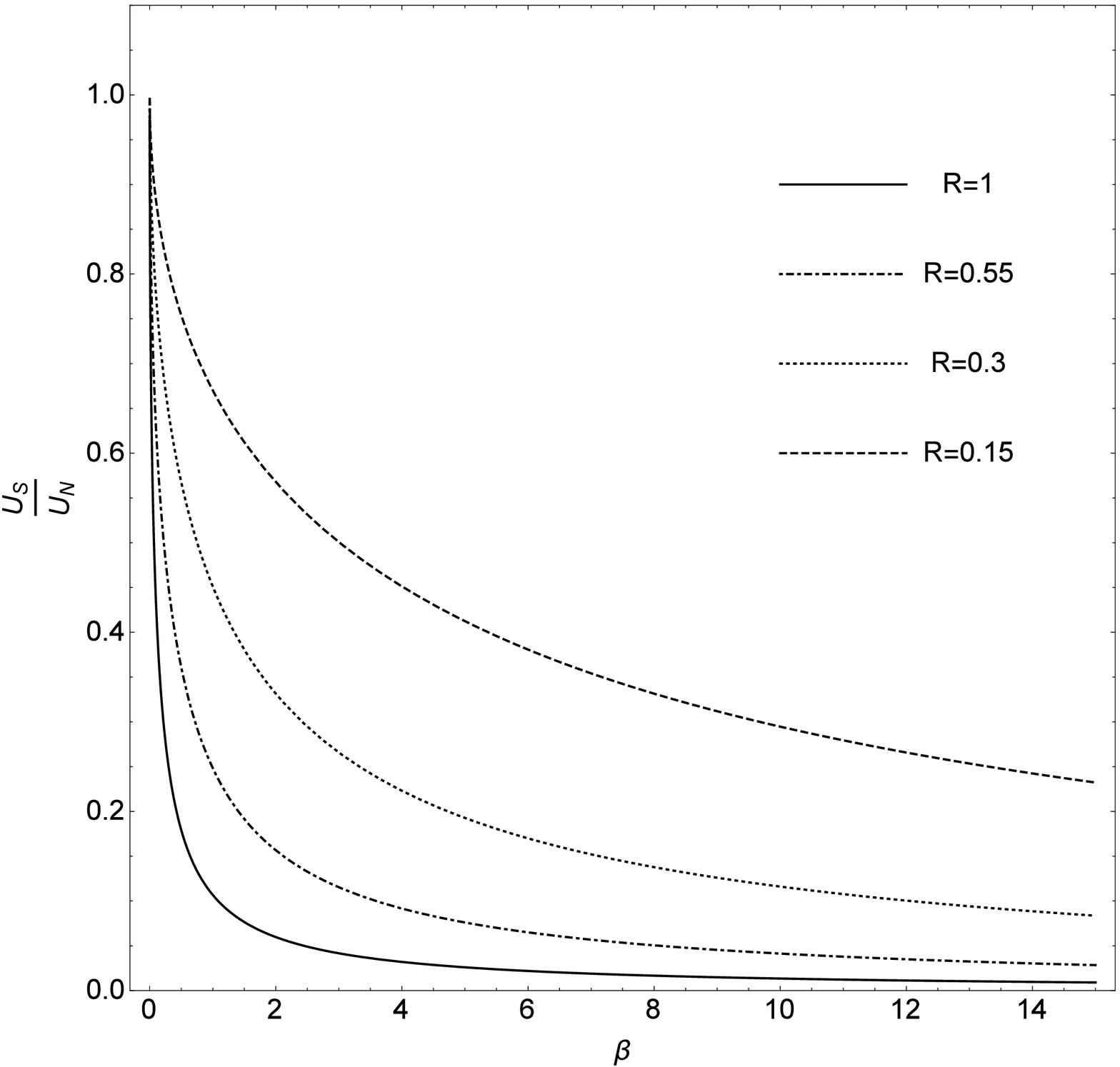}
		\includegraphics[scale=0.28]{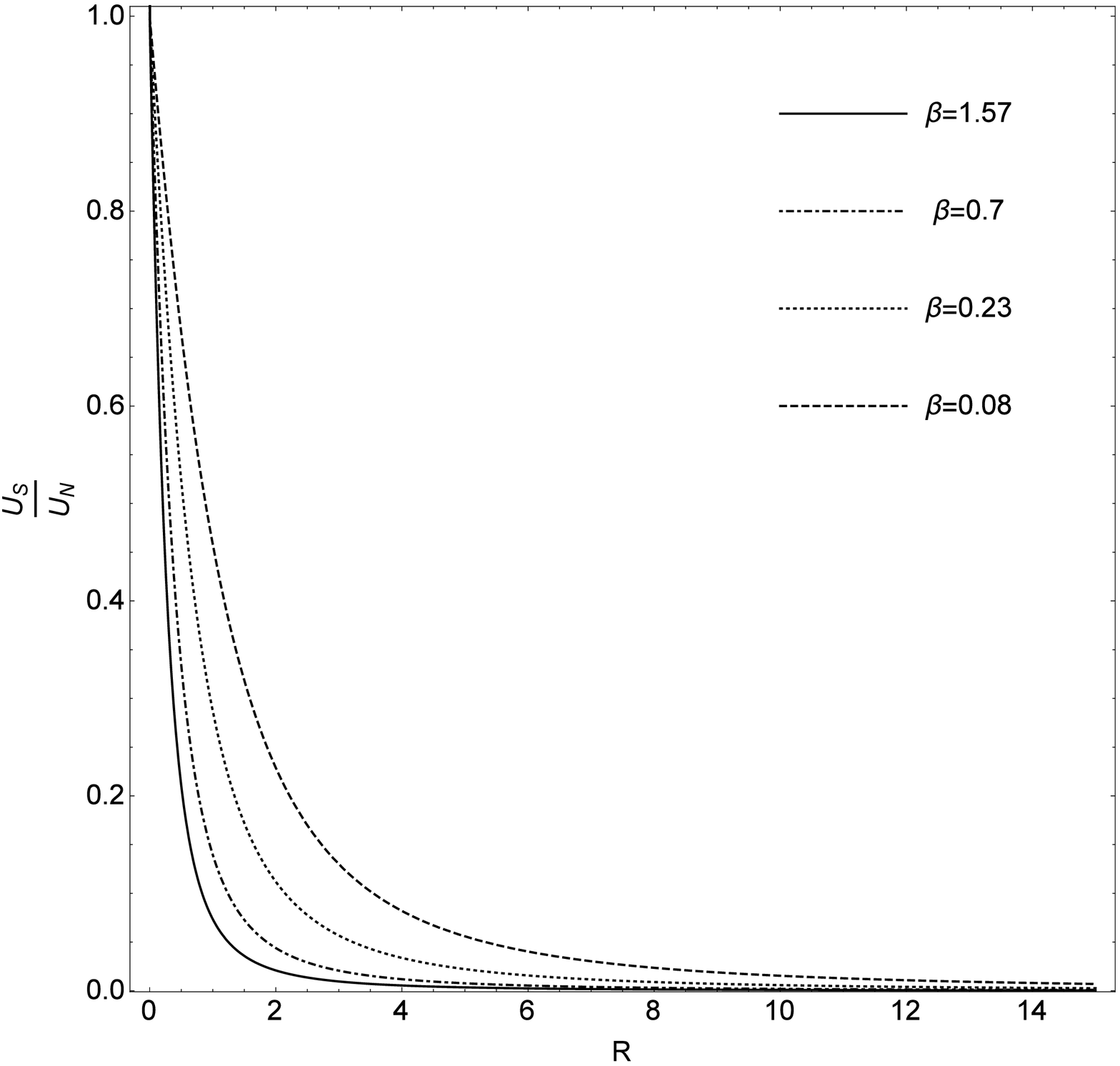}
		\caption{Upper panel: plot of the ratio $\frac{U_S}{U_N}$ as a
			function of $\beta$ parameter for different values of $R$. Lower
			panel: $\frac{U_S}{U_N}$ versus $R$ for some values of
			$\beta$.}\label{fig1}
	\end{center}
\end{figure}

As an example to verify the consequences of $\beta$, consider the
$4\beta=\frac{1}{R^2}$ case which leads to
$U_S\approx\frac{5}{18}U_N$, and thus the corresponding Jeans mass

\begin{eqnarray}\label{virial4}
&&M_S^J(\xi=1)=\left(\frac{18T}{\mu}\right)^{\frac{3}{2}}\left(\frac{3}{4\pi\rho_0}\right)^{\frac{1}{2}},
\end{eqnarray}

\noindent which is always bigger than $M^J$. It is worthwhile to
mention here that the Jeans mass may also be increased in the
framework of the Eddington-inspired Born-Infield gravity (De
Martino \& Capolupo 2017). In fact, at a constant radius $R$, the
Jeans mass is increased with increasing $\beta$. Mathematically,
as it is obvious from Fig.~\ref{fig1}, it is due to the fact that
the $\xi D$ term is negative and decreases with increasing
$\delta$ ($0<|\xi D|<1$) meaning that $U_S$ is always bigger than
$U_N$ for $\beta>0$. In this manner, since both $U_S$ and $U_N$
are negative, we have $0<\frac{U_S}{U_N}<1$.

In general, if we have $2\sqrt{\beta}=\frac{\zeta}{R}$, where
$\zeta$ is constant, then $\xi=\zeta=constant$, and we get the
following expression for the Jeans mass

\begin{eqnarray}\label{Jeansm}
&& M_S^J=\left(\frac{5T}{\mu(1+\zeta
	D)}\right)^{\frac{3}{2}}\left(\frac{3}{4\pi\rho_0}\right)^{\frac{1}{2}}>M^J.
\end{eqnarray}

%\section{Jeans instability in the R\'{e}nyi entropy formalism}

Now, consider a system consisting of $W$ discrete states while the
$i^{\textmd{th}}$ state can be occupied with probability $P_i$. It
has also been proposed that systems including long-range
interactions, such as gravity, may follow the power-law
probability distribution $P^q_i$ instead of the ordinary
distribution $P_i$, where $q$ is an unknown parameter evaluated by
using experiments and probably the other parts of physics
(R\'{e}nyi 1970; Masi 2005). Indeed, even an ideal gas can provide
the power-law probability distribution $P^q_i$ (Bir\'{o} 2013). In
this regard, the R\'{e}nyi entropy corresponding to the
probability distribution $P^q_i$ is defined as (R\'{e}nyi 1970;
Masi 2005)

\begin{eqnarray}\label{rs}
\mathcal{S}=\frac{1}{\delta}\ln\sum_{i=1}^{W} P_i^{1-\delta},
\end{eqnarray}

\noindent where $\delta\equiv1-q$, and the Shannon entropy is
recovered at the $\delta\rightarrow0$ limit (Masi 2005). This
entropy finally leads to (Moradpour et al 2018b)

\begin{eqnarray}\label{entr1}
\mathcal{S}=\frac{1}{\delta}\ln(1+\delta\frac{A}{4}),
\end{eqnarray}

\noindent and

\begin{eqnarray}\label{fr1}
\mathcal{F}=-\frac{Mm}{r^2}\frac{1}{\delta\pi r^2+1},
\end{eqnarray}

\noindent for the horizon entropy and the corresponding modified
NG, respectively (Komatsu 2017; Moradpour et al 2017; Moradpour et
al 2018b; Moradpour et al 2019). From mathematical point of view,
Eqs.~(\ref{entr1}) and~(\ref{fr1}) are obtainable replacing
$4\beta$ in Eqs.~(\ref{ent1}) and~(\ref{f1}) with $\delta\pi$,
respectively. One can easily check that this recipe is also
followable to find the modified Newtonian potential and the
corresponding gravitational potential energy from Eqs.~(\ref{p1})
and~(\ref{gpe1}), respectively.

%The only thing important is that, here, we are not
%allowed to use Eq.~(\ref{vir1}) based on ordinary probability
%distribution $P_i$ (Chung \& Hassanabadi 2019).

Although entropy, probability distribution and the phase space
geometries used to obtain Eqs.~(\ref{virial}) and~(\ref{vir1})
(Chung \& Hassanabadi 2019) differ from those based on
Eq.~(\ref{rs}) and the power-law probability distribution $P^q_i$
(Parvan \& Bir\'{o} 2005), it has been shown that the ideal gas
still satisfies Eq.~(\ref{vir1}) in the R\'{e}nyi formalism
(Parvan \& Bir\'{o} 2005). Thus, the results obtained by employing
EUP are also valid here, and one can get the exact from
of the outcomes by replacing $4\beta$ with $\delta\pi$. %Briefly,
%these consequences can confirm previous study showing that
%$\delta$ (and thus the existence of $q$ distribution) and $\beta$
%can be in close relation (Moradpour et al 2019).}

%%%%%%%%%%%%%%%%%%%%%%%%%%%%%%%%%%%%%%%%%%%%%%%%%%%%%%%%%%%%%%%%%%%%%%%%%%%%%%%%%%
\section{GUP and Jeans instability}

For $\beta=0$, Eq.~(\ref{gup}) leads to (Kempf et al. 1995)

\begin{eqnarray}\label{gupeta}
\Delta x\Delta p\geq\frac{1}{2}[1+\eta(\Delta p)^2],
\end{eqnarray}

\noindent which modifies the Newtonian potential corresponding to
mass $M$ as (Awada \& Ali 2014b)

\begin{eqnarray}\label{np1}
V_{GUP}(r)=-\frac{M}{r}\bigg[1+\frac{\Delta}{r^2}\big(\frac{2}{3}+\ln(4\pi
r^2)\big)\bigg],
\end{eqnarray}

\noindent where $\Delta=\frac{\eta}{144}$. Now, using
Eq.~(\ref{gpe}), one can get
\begin{figure}
	\begin{center}
		\includegraphics[scale=0.28]{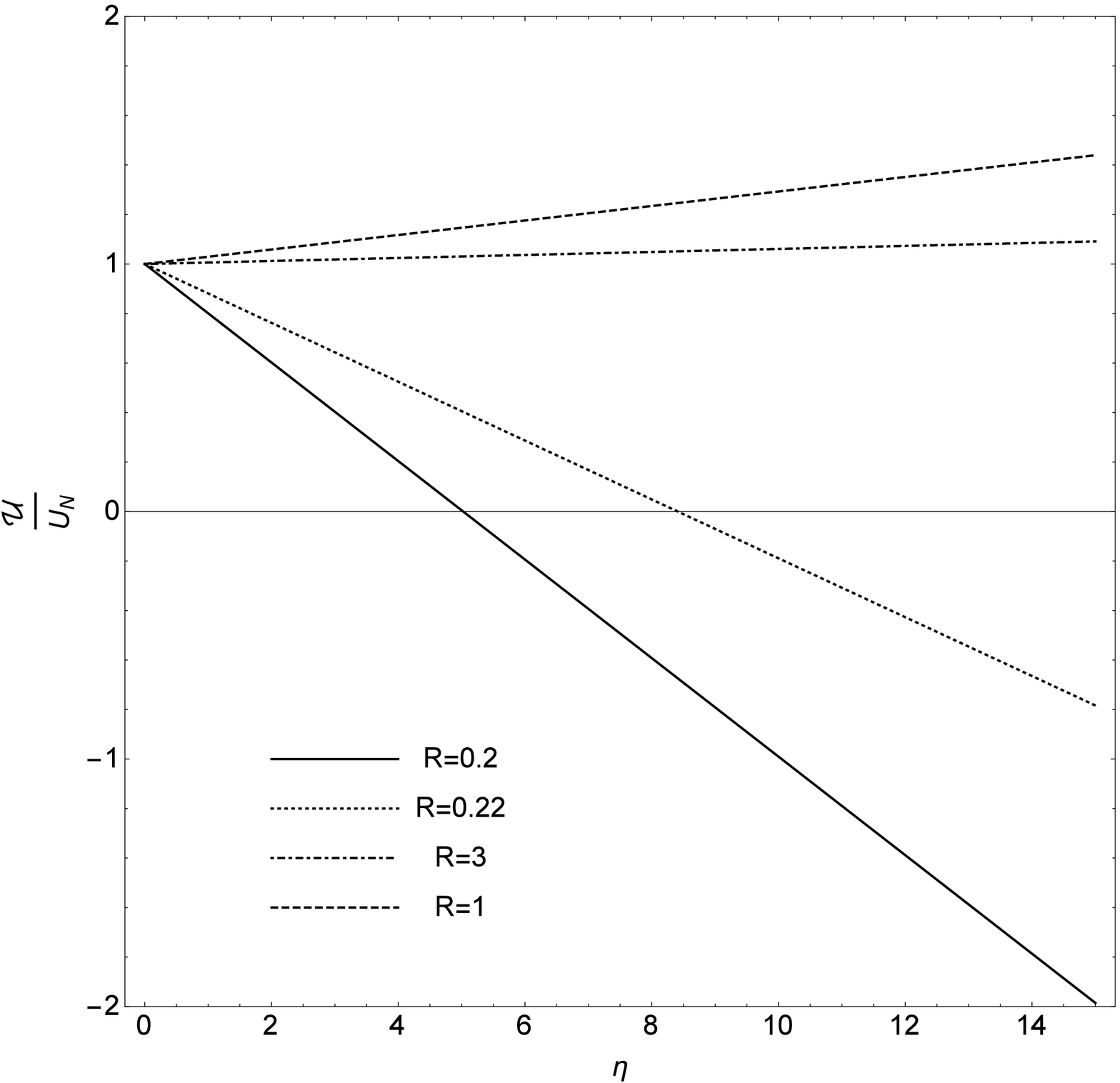}
		\includegraphics[scale=0.28]{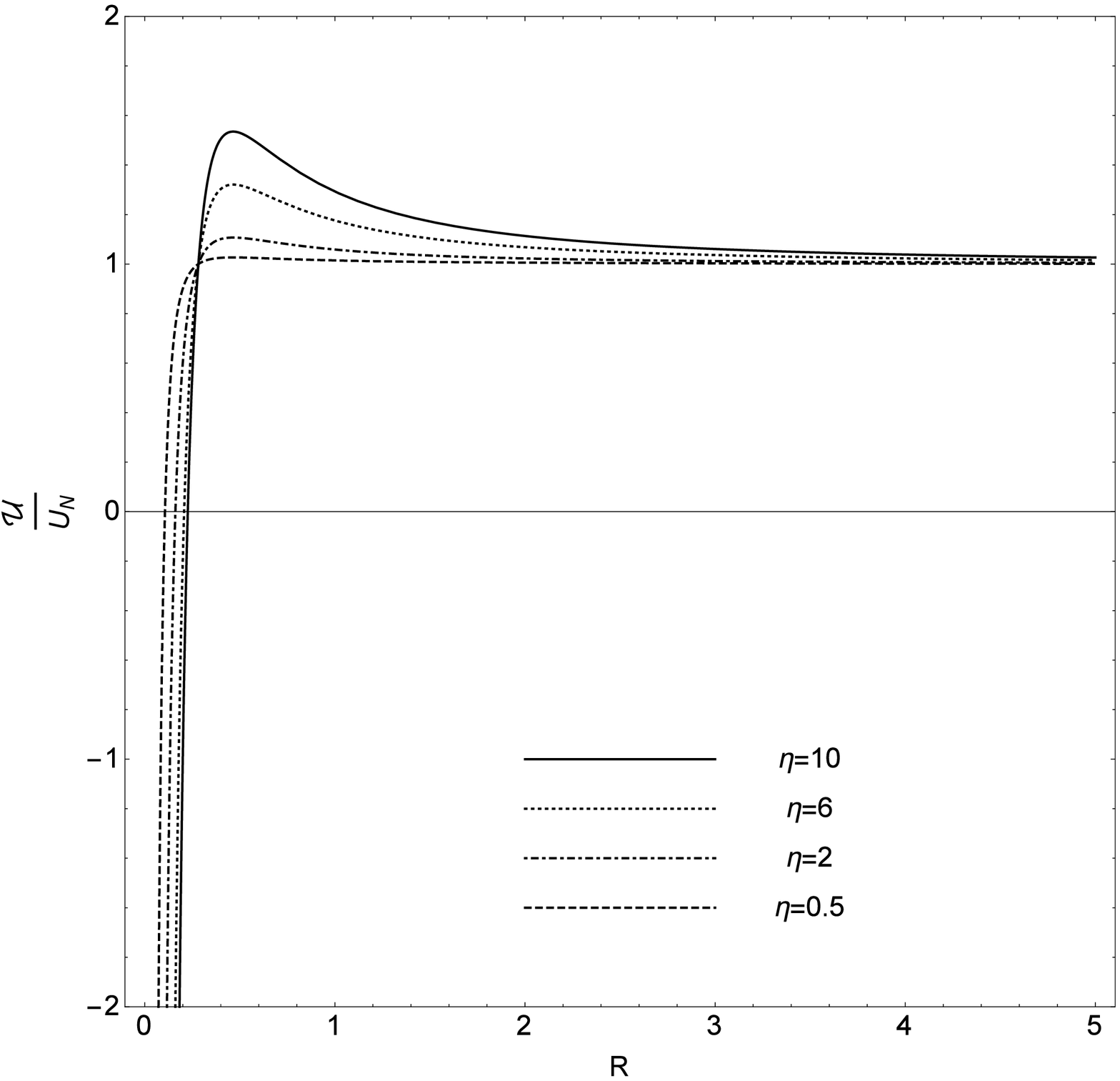}
		\caption{Upper panel: plot of the ratio $\frac{\mathcal{U}}{U_N}$
			as a function of $\eta$ parameter for different values of $R$.
			Lower panel: $\frac{\mathcal{U}}{U_N}$ versus $R$ for some values
			of $\eta$. Note that whenever $\mathcal{U}<U_N$, since both
			$\mathcal{U}$ and $U_N$ are negative, we have
			$\frac{\mathcal{U}}{U_N}>1$.}\label{fig2}
	\end{center}
\end{figure}

\begin{eqnarray}\label{gupp}
&&\mathcal{U}=\int_0^RV_{GUP}(r)dM=U_N+\Delta\mathcal{U}_c,\nonumber\\
&&\mathcal{U}_c=-\int_0^R\frac{M}{r^3}\big(\frac{2}{3}+\ln(4\pi
r^2)\big)dM=U_NB\nonumber,\\
&&B=\frac{5}{3R^2}\ln(4\pi R^2),
\end{eqnarray}

\noindent for the gravitational potential energy ($\mathcal{U}$)
corresponding to a cloud with mass $M$, density $\rho_0$ and
radius $R$. It has been shown that, up to the first order in
$\eta$, GUP modifies Eq.~(\ref{virial}) as (Motlaq \& Pedram 2014)

\begin{eqnarray}\label{vir2}
&&K=\frac{3}{2}NT-3\eta N\mu T^2,
\end{eqnarray}

\noindent where $\mu$ denotes again the mass of the ideal gas
particles. More studies on the thermodynamic properties of this
gas, such as its pressure, can be found in (Motlaq \& Pedram 2014;
Miraboutalebi \& Matin 2014). Bearing the Virial theorem in mind,
collapse will happen if the cloud mass $M$ satisfies

\begin{eqnarray}\label{vir3}
\!\!\!\!\!\!\!\frac{10T(\frac{1}{2\mu}-\eta
	T)}{(\frac{4\pi\rho_0}{3})^{\frac{1}{3}}}\leq
M^{\frac{2}{3}}\bigg[1+\frac{5\eta}{342(\frac{3M}{4\pi\rho_0})^{\frac{2}{3}}}\ln\big(4\pi(\frac{3M}{4\pi\rho_0})^{\frac{2}{3}}\big)\bigg].
\end{eqnarray}

\noindent Clearly, Eq.~(\ref{virial2}) is recovered at the
appropriate limit $\eta=0$. Moreover, since $\eta$ is finite
(Kempf et al. 1995; Nozari \& Fazlpour 2006; Nozari \& Etemadi
2012) and $M\gg1$, the above equation takes approximately the form

\begin{eqnarray}\label{vir4}
\frac{10T(\frac{1}{2\mu}-\eta
	T)}{(\frac{4\pi\rho_0}{3})^{\frac{1}{3}}}\leq M^{\frac{2}{3}}.
\end{eqnarray}

\noindent In this situation the Jeans mass ($\mathcal{M}^J$) is
obtained as

\begin{eqnarray}\label{vir5}
\mathcal{M}^J\approx M^J\big(1-2\eta\mu T\big)^{\frac{3}{2}},
\end{eqnarray}

\noindent meaning that the Jeans mass is decreased and is positive
only when $\eta\mu T<\frac{1}{2}$. In fact, depending on the value
of $\eta\mu T$, the ratio $\frac{\mathcal{M}^J}{M^J}$ can be
sensible.

While some Bok globules, as the interstellar clouds of dust and
gas, include star or at least experience the star formation
process (Bourke et al 1995; Kandori et al 2005; Launhardt et al
2000; Launhardt et al 2013; Vainio \& Vilja 2016; Yun and Clemens
1990), their mass is less than their corresponding Jeans mass
(Kandori et al 2005; Vainio \& Vilja 2016). This fact motivates
physicists to use modified gravity in order to revise Jeans mass
for describing the observations (Vainio \& Vilja 2016).

As an example, consider the Bok globule CB $188$ with mass $M_{CB\
	188}(\simeq7\cdot19\textmd{M}_\centerdot$) and
$M^J\simeq7\cdot7\textmd{M}_\centerdot$, where
$\textmd{M}_\centerdot$ denotes the Sun mass, which is at
temperature $T=19$ in Kelvin scale, and includes a protostar
(Vainio \& Vilja 2016). In this situation, assuming $M_{CB\
	188}=\mathcal{M}^J$, it is easy to see
$\frac{\ln(\mathcal{M}^J)}{(\mathcal{M}^J)^{\frac{2}{3}}}$ is
negligible and thus Eq.~(\ref{vir5}) is available which leads to

\begin{eqnarray}\label{eta}
\eta_{CB\ 188}\simeq\frac{0.0012}{\mu},
\end{eqnarray}

\noindent as a primary estimation for the value of $\eta$ in CB
$188$. In fact, an upper bound on $\mathcal{M}^J$ corresponds to a
lower bound on $\eta$, and for values bigger than $\eta_{CB\
	188}$, we have $\mathcal{M}^J<M_{CB\ 188}$. For a number of Bok
globules, including stars (Kandori et al 2005), following the
above recipe, the minimum values of $\eta$ have been addressed in
Table. 1. Here, we assumed that the host cloud is composed of $N$
identical particle with mass $\mu$. In reality, interstellar
clouds include a range of different particles, and therefore, the
mass of the particle with most abundance in the host cloud may be
considered as a primary estimation for $\mu$. In the case of CB
$188$, since it includes a protostar (Vainio \& Vilja 2016), we
follow Vainio \& Vilja (2016) and approximate $\mu$ with the
Hydrogen mass ($\sim
10^{9}~\textmd{eV}\sim10^{-19}~\textmd{M}_\textmd{P}$), where
$\textmd{M}_\textmd{P}$ denotes the Plank mass
($\textmd{M}_\textmd{P}\equiv\sqrt{\frac{\hbar c}{G}}=1$ in our
units, and we have $\textmd{M}_\textmd{P}\sim10^{-8}~\textmd{Kg}$
in the $\textmd{SI}$ units), which finally leads to $\eta_{CB\
	188}\sim10^{16}$. If we apply this assumption to all cases studied
in Table. 1, then the maximum value of $\eta$ is obtained for CB
$161$ as $\mu\sim10^{18}$ meaning that for values of $\eta$ bigger
than $10^{18}$, $\mathcal{M}^J$ of those Bok globules studied in
this paper is smaller than their mass.

It is also worthwhile to mention that, in the presence of the
equivalence principle and in the $\textmd{SI}$ units, we have
$\eta<2.3\times10^{60}$ (Feng et al 2017). Following Feng et al
(2017), simple calculations lead to $\eta<6.8\times10^{61}$ in our
units. The minimum upper bound on $\eta$ has also been reported as
$\eta<10^{21}$ (Das \& Vagenas 2008; Feng et al 2017). Finally, it
should be noted that a simple recipe to include the effect of the
mass of all ingredients of Bok globule is to model the Bok globule
as a system with $N$ particles of mass $\bar{\mu}$ such that $i$)
$N\equiv\sum_i N_i$, and $ii$) $\bar{\mu}\equiv\frac{\sum_i N_i
	\mu_i}{N}$, where $N_i$ and $\mu_i$ denote the number and mass of
the $i^{\textmd{th}}$ type particle, respectively. It means that,
since Hydrogen is the lightest atom, we have $\bar{\mu}>\mu$ if
there is something other than Hydrogen in the cloud, and
therefore, the obtained minimum values of $\eta$ will be decreased
by considering all ingredients of the Bok globules.

\begin{table}
	\caption{Approximation values for the minimum values of $\eta$ in
		some of the Bok globules by using the information of Vainio \&
		Vilja (2016). Here, $M$ and $T$ denote the mass and temperature of
		Bok globules, respectively.}
	\begin{tabular}{cccll}
		Bok globule & $T$ & $\frac{M}{\textmd{M}_\centerdot}$ & $\frac{M^J}{\textmd{M}_\centerdot}$ & $\eta\simeq\frac{1-(\frac{M}{M^J})^{\frac{2}{3}}}{2\mu T}$ \\ \hline\\
		CB $87$ & 11.4 & 2.73$\pm$0.24 & 9.6 & $\frac{0.0249}{\mu}$  \\ \hline\\
		CB $110$ & 21.8 & 7.21$\pm$1.64 & 8.5 & $\frac{0.0024}{\mu}$ \\ \hline\\
		CB $131$ & 25.1 & 7.83$\pm$2.35 & 8.1 &  $\frac{0.0004}{\mu}$ \\ \hline\\
		CB $161$ & 12.5 & 2.79$\pm$0.72 & 5.4 & $\frac{0.1424}{\mu}$  \\ \hline\\
		CB $184$ & 15.5 & 4.70$\pm$1.76 & 11.4 &  $\frac{0.0144}{\mu}$ \\ \hline\\
		FeSt $1-457$ & 10.9 & 1.12$\pm$0.23 & 1.4 & $\frac{0.0063}{\mu}$  \\ \hline\\
		Lynds $495$ & 12.6 & 2.95$\pm$0.77 & 6.6 & $\frac{0.0165}{\mu}$  \\
		\hline\\ Lynds $498$ & 11.0 & 1.42$\pm$0.16 & 5.7 &
		$\frac{0.0275}{\mu}$
	\end{tabular}
\end{table}

%At this
%step, the Hydrogen mass ($\approx10^9-\textmd{Mev}$) is also a
%suitable primary estimation for $\mu$.

%\textbf{As another example, consider the $2\eta\mu T\ll1$
%situation for which Eq.~(\ref{vir2}) leads to
%$K\approx\frac{3}{2}NT$ (Motlaq \& Pedram 2014).} In reality,
%$R\gg1$ (Longair 1998) meaning that one may approximately write
%$\mathcal{U}_c\approx\frac{10}{3R}U_N$ up to the first order in
%$\frac{1}{R}$ leading to $\mathcal{U}\approx
%U_N(1+\frac{\eta}{216R})$, and
%
%\begin{eqnarray}\label{virial5}
%\mathcal{M}^J=\frac{1}{8}\big[\sqrt{4(M^J)^{\frac{2}{3}}+(\eta^\prime)^2}-\eta^\prime\big]^3,
%\end{eqnarray}
%
%\noindent where
%$\eta^\prime=\frac{\eta}{216}(\frac{4\pi\rho_0}{3})^{\frac{1}{3}}$,
%for the Jeans mass. Defining
%$\lambda=\frac{\eta}{216}(\frac{5T}{\mu})$, Eq.~(\ref{virial5})
%can be rewritten as
%
%\begin{eqnarray}\label{virial6}
%\mathcal{M}^J=M^J\big[\sqrt{1+(\frac{\lambda}{2M^J})^2}-\frac{\lambda}{2M^J}\big]^3,
%\end{eqnarray}
%
%\noindent which is always smaller than $M^J$ whenever
%$\frac{\lambda}{2M^J}>0$.

In fact, since $\eta$ is positive (Kempf et al. 1995; Motlaq \&
Pedram 2014; Nozari \& Fazlpour 2006), we have $\Delta B>0$ if
$R^2>\frac{1}{4\pi}$ which leads to $\mathcal{U}<U_N$ meaning that
the Jeans mass is decreased in this situation (see
Fig.~\ref{fig2}). Moreover and in agreement with Fig.~\ref{fig2},
although the deviation of $\frac{\mathcal{U}}{U_N}$ from $1$ is
sensible for $R<3$, depending on the values of temperature, $\eta$
and $\mu$, the $\frac{\mathcal{M}^J}{M^J}$ quantity may be
appreciable. It is due to this fact that while
$\frac{\mathcal{U}}{U_N}$ is independent of temperature, Jeans
mass depends on it.
%%%%%%%%%%%%%%%%%%%%%%%%%%%%%%%%%%%%%%%%%%%%%%%%%%%%%%%%%%%%%%%%%%%%%
\section{Summary}

The implications of GUP and EUP on the Jeans mass have been
studied. The results indicate that, compared with the Newtonian
regime, the gravitational potential energy of the cloud is
decreased when EUP is valid, leading to growth in the Jeans mass.
It has been argued that since gravity is a long-range interaction,
one may use the R\'{e}nyi entropy formalism to study the
gravitational systems (Komatsu 2017; Moradpour et al 2017;
Moradpour et al 2018a; Moradpour et al 2018b; Moradpour et al.
2019b). This entropy~(\ref{entr1}) is also confirmed by EUP
(Moradpour et al 2019), and leads to Eq.~(\ref{fr1}) (Moradpour et
al 2018b). Additionally, as we addressed, Eq.~(\ref{vir1}) is
valid in both the EUP and R\'{e}nyi frameworks. Hence, the EUP
results, obtained in this paper, are also available in the
framework of R\'{e}nyi entropy (one just needs to replace $4\beta$
in the results of second section with $\delta\pi$ to obtain the
corresponding results in the R\'{e}nyi formalism).

Moreover, we found out that GUP can increase the gravitational
potential energy which reduces the Jeans mass. Therefore, GUP
suggests that gravitational collapse can happen for clouds with
mass smaller than the ordinary Jeans mass obtained from NG
($M^J$). The Bok globules observations confirm the existence of
collapse process in clouds whose their mass is smaller than $M^J$
(Kandori et al 2005; Launhardt et al 2000; Vainio \& Vilja 2016).
Fitting such observations to the modified Jeans mass obtained from
GUP, one may find some primary estimations for $\eta$ (For example
Eq.~(\ref{eta})). The evolution of fragments and perturbations (in
both astrophysical and cosmic scales) under the shadow of the GUP
modifications to Jeans mass are also another interesting subjects,
which are out of the scope of this project, and are followable in
the subsequent works.
%%%%%%%%%%%%%%%%%%%%%%%%%%%%%%%%%%%%%%%%%%%%%%%%%%%%%%
\section*{Acknowledgment}
We are grateful to the anonymous reviewer for worthy comments. The
work of H. Moradpour has been supported financially by Research
Institute for Astronomy \& Astrophysics of Maragha (RIAAM) under
research project No. $1/6025-55$.
%%%%%%%%%%%%%%%%%%%%%%%%%%%%%%%%%%%%%%%%%%%%%%%%%%%%%%%%%%%%%%%%%%%%%%%%%%%%%%%%%%%%%%%%

\label{lastpage}
\end{document}